\documentclass[graybox]{svmult}
\usepackage{mathptmx}                        
\usepackage{helvet}                                                 
\usepackage{courier}                                                
\usepackage[T1]{fontenc}                                                
\usepackage[utf8]{inputenc}                                              
\usepackage{url}     
\usepackage[UKenglish]{babel}
\usepackage{amsmath}
\usepackage{amssymb}
\usepackage{amstext}
\usepackage{dsfont}
\usepackage[decimalsymbol=.]{siunitx}
\usepackage{xfrac}
\usepackage{graphicx}
\usepackage{tabularx}
\usepackage{makeidx}                                   
\usepackage{multicol}                            
\makeindex       
\usepackage[bottom]{footmisc}                  
\newcommand{\Reynolds}{Re}
\newcommand{\Womersley}{W\!o}
\begin{document}
%
%
\title*{On phase asymmetries in oscillatory pipe flow} 
\titlerunning{On phase asymmetries in oscillatory pipe flow} 
%
\author{Daniel~Feldmann and Claus~Wagner}                           
\authorrunning{D.~Feldmann \& C.~Wagner}                   
%
\institute{Daniel Feldmann
\at Institute of Aerodynamics and Flow Technology, German Aerospace Center, Bunsenstr. 10, 37073 G\"ottingen, Germany, 
\email{feldmann@mailbox.org}, ORCID: 0000-0002-6585-2875
\and Claus Wagner
\at Institute of Thermo- and Fluid Dynamics, Technical University Ilmenau, Helmholtzring 1, 98693 Ilmenau, Germany
\email{claus.wagner@tu-ilmenau.de}
}
%
\maketitle
%
%
%

%
%
%
\abstract{
We present results from direct numerical simulations (DNS) of oscillatory pipe
flow at several dimensionless frequencies $\Womersley\in\{6.5,13,26\}$ and one
fixed shear Reynolds number $\Reynolds_{\tau}=\num{1440}$.
Starting from a fully-developed turbulent velocity field at that
$\Reynolds_{\tau}$, the oscillatory flow either relaminarises or reaches a
conditionally turbulent or strongly asymmetric state depending on $\Womersley$.
The numerical method is validated by demonstrating excellent agreement of our
DNS results with experimental data and analytical predictions from literature
for the limiting cases of non-oscillating but turbulent and oscillating but
laminar pipe flow.
For an oscillating turbulent pipe flow we further found a very good agreement
between qualitative descriptions of the characteristic flow features observed in
experiments and our DNS.
Here, we focus on the observation of a strongly asymmetric behaviour between the
positive and the negative half-cycles of the oscillatory pipe flow at
$\Womersley=\num{6.5}$.
}
%
%
%
%
%
%
%
\section{Introduction}
\label{sec:introduction}
\begin{figure}[htbp]
\sidecaption
\includegraphics[width=0.60\textwidth]{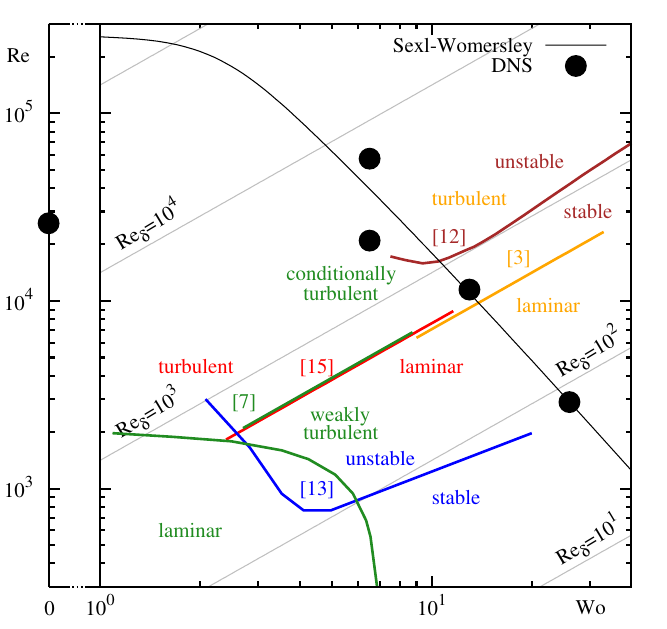}
\caption{Parameter space in terms of the Womersley number ($\Womersley$) and the
peak Reynolds number ($\Reynolds$) characterising oscillatory pipe flow.
Partly ambiguous findings from experimental investigations, i.e.
\cite{hino1976, eckmann1991, zhao1996}, as well as theoretical results from
linear stability analyses \cite{trukenmueller2006, thomas2012} are also
illustrated to indicate the regimes of occurrence of laminar and turbulent
(or at least non-laminar) states in the context of the oscillatory pipe flow
scenario. Analytical relation according to laminar theory (SW) for
$\Reynolds_{\tau}=\num{1440}$ as reference.}
\label{fig:woReRange}
\end{figure}

%
For many industrial applications and especially for bio-fluid dynamics, e.g.
respiratory airflow (oscillatory) or vascular blood flow (pulsatile), decay and
amplification of turbulence play an important role.
The onset of turbulence in such wall-bounded time-periodic fluid motions
considerably affect the mixing and transport efficiency and can rapidly change
shear forces acting on the system's wall.
Contrarily to the onset of turbulence in statistically steady pipe flows, see
e.g. Avila et al. \cite{avila2011}, turbulence in statistically unsteady pipe
flows is far from understood.
Some of the few existing theoretical, numerical and experimental studies on
transition to turbulence in such flows are summarised and discussed in e.g.
\cite{feldmann2012, feldmann2014}.
The most important results are also collated in fig.~\ref{fig:woReRange}.
%
%
To our knowledge, the existing numerical studies only consider oscillatory flows
over a flat plate or in two-dimensional channels and cover rather few and low
Womersley numbers $\Womersley$, or focus on pulsatile rather than oscillatory
flows.
To complement the findings summarised in fig.~\ref{fig:woReRange}, we perform
three-dimensional direct numerical simulations (DNS) of oscillatory pipe.
In the following we report and discuss the vastly asymmetric behaviour of an
oscillatory pipe flow at $\Womersley=\num{6.5}$.
%
%
%

%
%
%
\section{Numerical method}
\label{sec:numericalMethod}
%
We consider an incompressible Newtonian fluid with constant properties confined
by a cylindrical wall of diameter $D$ and length $L$.
The governing equations read
%
\begin{equation}
\nabla\cdot\vec{u}=0
\hspace{3.5mm}
\text{and}
\hspace{3.5mm}
\frac{\partial\vec{u}}{\partial t} +
\left(\vec{u}\cdot\nabla\right)\vec{u} +
\nabla p^{\prime} -
\frac{1}{\Reynolds_{\tau}}\nabla^{2}\vec{u} =
-\vec{P}(t)
\hspace{0.5mm} \text{,}
\label{eq:navierStokesEquations}
\end{equation}
%
where the flow is driven in axial direction $z$ by the prescribed mean pressure
gradient
%
\begin{equation}
P_{z}(t) = \frac{\partial\langle p \rangle}{\partial z} \equiv
-4\cos\!\left(\frac{4\Womersley^{2}t}{\Reynolds_{\tau}}\right)\!
\text{,}\hspace{3.00mm}
P_{\varphi} \equiv 0
\text{,}\hspace{3.00mm}
P_{r} \equiv 0
\hspace{3.0mm}\text{with}\hspace{3.0mm}
p = p^{\prime} + \langle p \rangle
\hspace{0.25mm}\text{.}
\label{eq:meanPressureGradient}
\end{equation}
%
The time is denoted by $t$ and $\vec{u}$ is the velocity vector with its
components $u_{z}$, $u_{\varphi}$, and $u_{r}$ pointing along the cylindrical
coordinates in axial ($z$), azimuthal ($\varphi$) and radial ($r$) direction.
According to the Reynolds decomposition the pressure $p$ is separated in its
fluctuating part $p^{\prime}$ and its mean part $\langle p \rangle$, such that
periodic boundary conditions can be employed for $p^{\prime}$ in $z$ and in
$\varphi$.
The control parameters and the normalisation in
eqs.~(\ref{eq:navierStokesEquations}) and (\ref{eq:meanPressureGradient}) are
given by the Womersley number
$\Womersley = \sfrac{D}{2}\sqrt{\sfrac{\omega}{\nu}}$ and the shear
Reynolds number $\Reynolds_{\tau}=\sfrac{u_{\tau}D}{\nu}$.
Here, $\omega=\sfrac{2\pi}{T}$ is the oscillation frequency, $u_{\tau}$ is the
friction velocity of a fully-developed turbulent pipe flow at $\Womersley=0$,
and $\nu$ is the kinematic viscosity.
The oscillatory pipe flow is fully characterised by $\Womersley$ and the peak
Reynolds number $\Reynolds = \sfrac{u_{\text{p}} D}{\nu}$, where
$u_{\text{p}} = \max\:u_{\text{b}}(t)$ is the peak velocity within
$0\leqslant t<T$ and $u_{\text{b}}$ is the instantaneous bulk velocity.
Eqs.~(\ref{eq:navierStokesEquations}) and~(\ref{eq:meanPressureGradient}) are
complemented by periodic boundary conditions in $z$ and in $\varphi$ and with
no-slip conditions at $r=\sfrac{D}{2}$.
They are discretised using a fourth order accurate finite volume method on
staggered grids, based on Schumann's \cite{schumann1973} volume balance
procedure.
Time advancement is done using a leapfrog-Euler time integration scheme where
the time step is adapted using a von Neuman stability criterion.
Further details on the numerical method are given in
\cite{schmitt1986, shishkina2007, feldmann2012}.
%
%
%
%
%
%
\section{Fully turbulent test case ($\Womersley=0$)}
\label{subsec:fullyTurbulentTestCase}
\begin{figure}[htbp]
\centering
\includegraphics[scale=1.0]{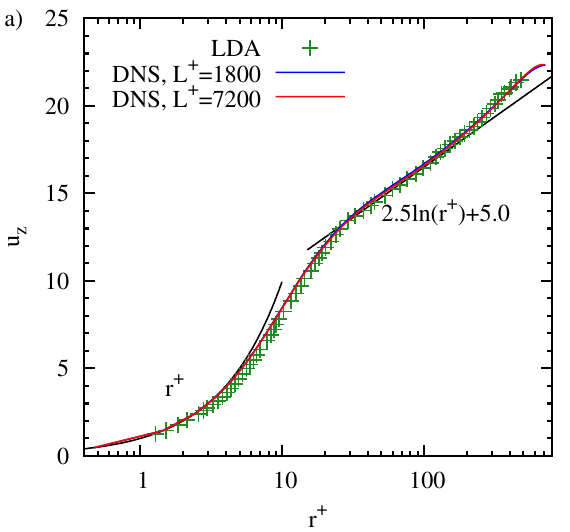}
\includegraphics[scale=1.0]{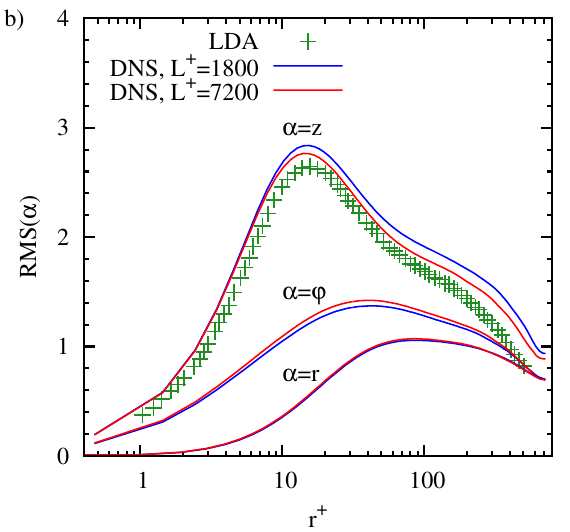}
\caption[Radial profiles for $\Womersley=\num{0}$]
{Radial profiles of the mean axial velocity component
($\langle u_{z}\rangle_{t,z,\varphi}$) and of the RMS velocity fluctuations
($\sqrt{\langle u^{\prime 2}_{\alpha}\rangle_{t,z,\varphi}}$)
as obtained by DNS for $\Reynolds=25900$ compared to experimental results from
Durst et al. \cite{durst1996} as obtained by LDA for
$\Reynolds=18500$.}
\label{fig:velocityAndRmsProfilesDurst}
\end{figure}

%
To provide adequate initial conditions for all oscillating pipe flow DNS and to
demonstrate the capability of the used numerical method to properly reproduce
the physics of turbulent wall-bounded shear flows, we performed DNS of a
non-oscillating but fully-developed turbulent pipe flow at
$\Reynolds_{\tau}=\num{1440}$.
The resulting bulk Reynolds number of the statistically-steady turbulent pipe
flow is $\Reynolds=\num{25960}$.
The radial profile of the mean axial velocity is shown in
fig.~\ref{fig:velocityAndRmsProfilesDurst}a).
%
%
There is an excellent agreement with the linear law within the viscous sub-layer
($r^{+}<5$) close to the pipe wall as well as with the logarithmic law within
the inertial sub-layer ($30<r^{+}<100$).
Furthermore, there is a good agreement between our DNS results and experimental
data obtained by Durst et al. \cite{durst1996} using laser-Doppler anemometry
(LDA) in a pipe flow at a slightly lower $\Reynolds=\num{18500}$.
Fig.~\ref{fig:velocityAndRmsProfilesDurst}b) shows all three components of the
RMS velocity fluctuations.
There is a fairly good overall agreement with the experimental results,
despite the lower $\Reynolds$ of the LDA data.
The used spatial resolution expressed in wall units ($\sfrac{\nu}{u_{\tau}}$),
denoted by $^+$, is given in tab.~\ref{tab:dnsSetUp}.
%
\begin{table}[htbp]
\centering
\caption[DNS parameters]
{Resulting flow state, characteristic parameters, computational domain length
and spatial resolution (given in wall units denoted by ${}^+$) of the performed pipe flow
simulations at $\Reynolds_{\tau}=\num{1440}$.}
\label{tab:dnsSetUp}
\begin{tabularx}{1.00\textwidth}{lXSXSXSXS}
\hline\noalign{\smallskip}
{Flow state} &
{} &
{$\Womersley$} &
{} &
{$\Reynolds$} &
{} &
{$L^{+}$} &
{} &
{$\Delta z^{+} \times r^{+}\Delta \varphi \times \Delta r^{+}$}\\
\noalign{\smallskip}\svhline\noalign{\smallskip}
{Fully turb.} && {$0$}   && {$25960$}           && {$1800$, $7200$} && {$7.0\times(0.03-4.4)\times(0.5-11.1)$}\\
{Fully turb.} && {$0$}   && {$25960$}           && {$1700$, $5100$} && {$6.7\times(0.02-4.4)\times(0.5- 6.7)$}\\
{Asymmetric}  && {$6.5$} && {$21000$ / $57370$} && {$7200$}         && {$7.0\times(0.03-4.4)\times(0.5-11.1)$}\\
{Asymmetric}  && {$6.5$} && {$21000$ / $57370$} && {$1700$, $5100$} && {$6.7\times(0.02-4.4)\times(0.5- 6.7)$}\\
{Cond. turb.} && {$13$}  && {$11510$}           && {$1800$}         && {$7.0\times(0.03-4.4)\times(0.5-11.1)$}\\
{Cond. turb.} && {$13$}  && {$11510$}           && {$1700$, $5100$} && {$6.7\times(0.02-4.4)\times(0.5- 6.7)$}\\
{Laminar}     && {$26$}  && {$2910$}            && {$1800$, $7200$} && {$7.0\times(0.03-4.4)\times(0.5-11.1)$}\\
\noalign{\smallskip}\hline\noalign{\smallskip}
\end{tabularx}
\end{table}
%
As discussed in \cite{feldmann2012}, the pipe domain of length $L^{+}=1800$ is
sufficiently long and the spatial resolution is sufficiently fine to capture
all relevant length scales in the turbulent pipe flow.
For all simulations the adaptive time step varies in the range
$\num{6e-6}<\Delta t<\num{3.9e-5}$ and is thus much smaller than the Kolmogorov
time scales in the fully-developed turbulent pipe flow.
%
%
%
\section{Laminar oscillating test case ($\Womersley=\num{26}$)}
\label{subsec:laminarOscillatingTestCase}
\begin{figure}[htbp]
\sidecaption
\centering
\includegraphics[width=0.60\textwidth]{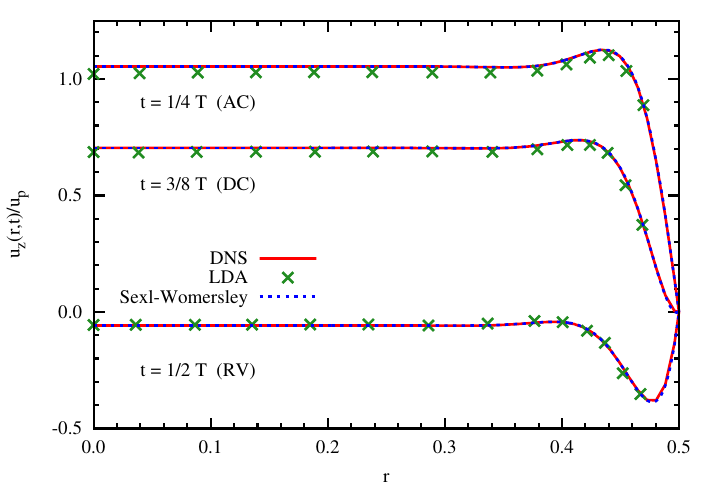}
\caption[Radial profiles for $\Womersley=\num{26}$]
{Radial profiles of the instantaneous axial velocity component
$u_{z}(r,t)$ at several oscillation phases as obtained by DNS for
$\Womersley=\num{26}$ in comparison to predictions from analytical theory (SW)
according to eq. (\ref{eq:sexlWomersleyVelocity}) and experimental data measured
by Eckmann \& Grotberg \cite{eckmann1991} using LDA in an oscillating laminar
pipe flow at a comparable setup at $\Womersley=\num{26.5}$ and
$\Reynolds=\num{8430}$.}
\label{fig:velocityProfilesEckmann}
\end{figure}

%
For a laminar, axially symmetric and fully-developed oscillatory pipe flow there
exists a closed-form solution to eqs. (\ref{eq:navierStokesEquations}) and
(\ref{eq:meanPressureGradient}) for the axial velocity component, hereinafter
referred to as Sexl-Womersley (SW) solution \cite{sexl1930, womersley1955}.
It reads
%
\begin{equation}
u_{z,\text{SW}}(r,t) = \Re \left\{
-\imag\frac{\Reynolds_{\tau}}{\Womersley^{2}}
\left[1 - \frac{J_{0}\left(2\imag^{\frac{3}{2}} \Womersley r\right)}
               {J_{0}\left( \imag^{\frac{3}{2}} \Womersley   \right)}\right]
e^{ \imag 4 \Womersley^{2} / \Reynolds_{\tau} t} \right\}
\label{eq:sexlWomersleyVelocity}
\end{equation}
%
using our normalisation, where $J_{0}$ is Bessel's function of first kind and
zeroth order and $\imag$ is the imaginary unit.
For $\Womersley\rightarrow\num{0}$ the SW solution takes the form of a
quasi-steady version of the Hagen-Poiseuille solution for laminar pipe flow with
a parabolic distribution.
For increasing $\Womersley$, the flow gradually becomes unsteady.
The parabolic velocity profile disappears in advantage of a plateau like
velocity distribution in the central region of the pipe, where inertial forces
now become dominant.
The velocity maximum occurs in combination with inflection points in the
vicinity of the pipe wall rather than at the pipe centre line.
In this near-wall region, i.e. the Stokes layer (of thickness $\delta$), the
viscous forces are still dominant.
During phases of bulk flow reversal (RV) a phenomenon called coaxial counterflow
takes place within the Stokes layer, when the viscous near-wall flow and the
inviscid core flow are directed in opposite ways at the same time.
All these features, which are typical for SW flow are exemplarily shown in
fig.~\ref{fig:velocityProfilesEckmann} for a moderately high Womersley number
of $\Womersley=26$.
%
%
\par
We performed a DNS of oscillatory pipe flow at $\Reynolds_{\tau}=\num{1440}$ and
$\Womersley=\num{26}$ using the fully-developed turbulent flow field described
in sec.~\ref{subsec:fullyTurbulentTestCase} as initial condition.
At $t=\num{0}$ the driving pressure gradient starts to vary between \num{-4} and
\num{4} in a single harmonic manner  with a period of
$T=\sfrac{\pi\Reynolds_{\tau}}{2\Womersley^{2}}=\num{3.3}$.
As a consequence the mean flow starts to decelerate; the turbulence intensity
decreases likewise.
We found that the flow field laminarises entirely within the first \num{20}
oscillation cycles.
However, to converge to a fully symmetric oscillatory pipe flow as described by
eq. (\ref{eq:sexlWomersleyVelocity}) it takes the flow at least another \num{50}
periods due to inertia.
Fig. \ref{fig:velocityProfilesEckmann} reflects an excellent overall agreement
between radial profiles of $u_{z}$ extracted from our DNS results for
$t>\num{70}T$, predictions from laminar theory (SW), and experimental results
obtained by Eckmann \& Grotberg \cite{eckmann1991} using LDA.
At this stage, the peak Reynolds number takes the value $\Reynolds=\num{2910}$.
As shown in fig. \ref{fig:woReRange} this result is also consistent with
$\Reynolds$ according to the SW solution, which we have obtained by integrating
eq. (\ref{eq:sexlWomersleyVelocity}) over the pipe cross-sectional area and
solving the resulting expression for the maximum value of the bulk velocity
within one cycle.
Repeating this for several $\Womersley$ enables us to determine an analytical
relation between $\Womersley$ and the resulting $\Reynolds$ for a given
$\Reynolds_{\tau}=1440$, which is also depicted in fig. \ref{fig:woReRange}.
Despite the relatively high $\Reynolds$, the well-correlated initial turbulence
is not sustained in the resulting oscillatory flow and our numerical study thus
supports the experimental results by Eckmann \& Grotberg \cite{eckmann1991} who
found oscillatory pipe flow to be laminar at this
$\Reynolds_{\delta}=\sqrt{\sfrac{1}{2}}\sfrac{\Reynolds}{\Womersley}=\num{79}$.
High-frequency time series of $p$ and $\vec{u}$ (not shown here) do not reflect
any fluctuations over the last \num{50} cycles and thus we are not able to
confirm the small near-wall instabilities prognosticated by Trukenm\"uller
\cite{trukenmueller2006} from quasi-steady linear stability analysis.
%
%
%
\section{Conditionally turbulent case ($\Womersley=\num{13}$)}
\label{sec:conditionallyTurbulentCase}
%
For $\Womersley=\num{13}$ the oscillation period is increased by a factor of
four ($T=\num{13.4}$) and so is the computational effort to simulate one cycle
with this halved $\Womersley$ and the same shear Reynolds number.
When compared to the case with $\Womersley=\num{26}$ the flow converges within
much fewer cycles to a pure oscillatory state, as reported earlier
\cite{feldmann2012, feldmann2014}.
This faster relaxation from the non-oscillating initial condition can be mainly
explained by two mechanisms.
First, the smaller $\Womersley$ implies a longer phase of deceleration (DC);
i.e. the driving force acts for a longer period of time against the initial bulk
flow direction.
Second, the flow does not become laminar.
Turbulent mixing enhances the deceleration of the bulk flow by intensified
wall-normal transport of kinetic energy from the high-momentum fluid in the core
region to the near-wall Stokes layer, where viscous forces dominate.
After the flow has reached an oscillatory state ($t>\num{27}$), the final peak
Reynolds number results in $\Reynolds=\num{11510}\pm\num{480}$ and thus
$\Reynolds_{\delta}=\num{626}$.
Despite the flow being intermittently turbulent, this value is surprisingly
close to $\Reynolds=11010$ predicted by the SW solution for a completely laminar
flow, see fig. \ref{fig:woReRange}.
\par
\begin{figure}[htbp]
\includegraphics[scale=1.00]{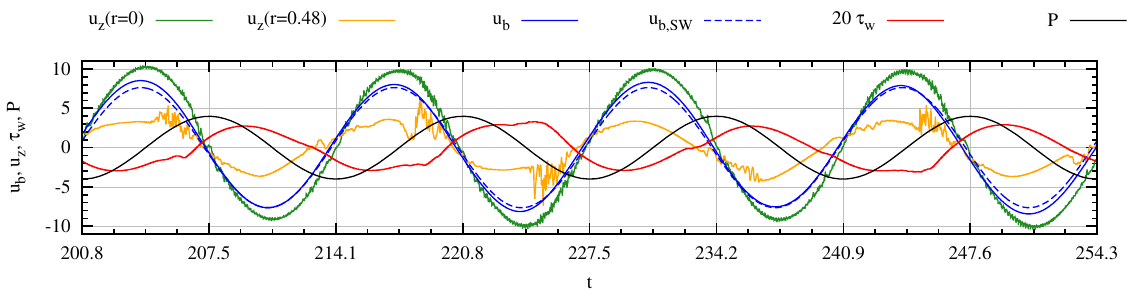}
\caption[Time series for $\Womersley=\num{13}$]
{Time series of the axial velocity, the bulk flow, and the wall-shear
stress as predicted by DNS in a pipe domain of length $L^{+}=\num{1800}$.
The flow is driven by a prescribed pressure gradient $P(t)$ for $\Womersley=13$.
Analytical predictions according to laminar theory (SW) are shown as reference.}
\label{fig:timeLinesConditionallyTurbulent}
\end{figure}

Fig. \ref{fig:timeLinesConditionallyTurbulent} presents time series of $u_{z}$
to illustrate the conditionally turbulent character of this oscillatory pipe
flow.
%
%
It can be seen that small velocity fluctuations are abruptly amplified near
the wall during the early stage of DC, e.g. $\num{205}<t<\num{207}$.
During the bulk flow reversal (RV), e.g. $t\approx200$, these fluctuations are
damped again and they further decay during the following acceleration phase
(AC); at least close to the wall.
This asymmetry between AC and DC is also reflected by the time signal of the
wall-shear stress plotted in fig.~\ref{fig:timeLinesConditionallyTurbulent}.
All these observations are in very good qualitative agreement with observations
from experimental investigations by e.g. Eckmann \& Grotberg
\cite{eckmann1991}, who report on turbulent bursts in the vicinity of the wall
during DC and a resulting asymmetry in each half-cycle for oscillatory pipe
flows at $\Reynolds_{\delta}>\num{500}$.
%
%
%
\section{Asymmetric case ($\Womersley=\num{6.5}$)}
\label{sec:asymmetricCase}
\begin{figure}[htbp]
\centering
\includegraphics[scale=1.00]{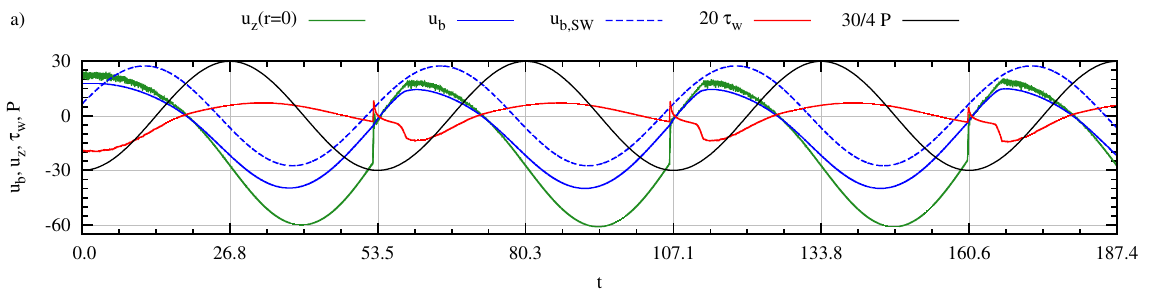}
\includegraphics[scale=1.00]{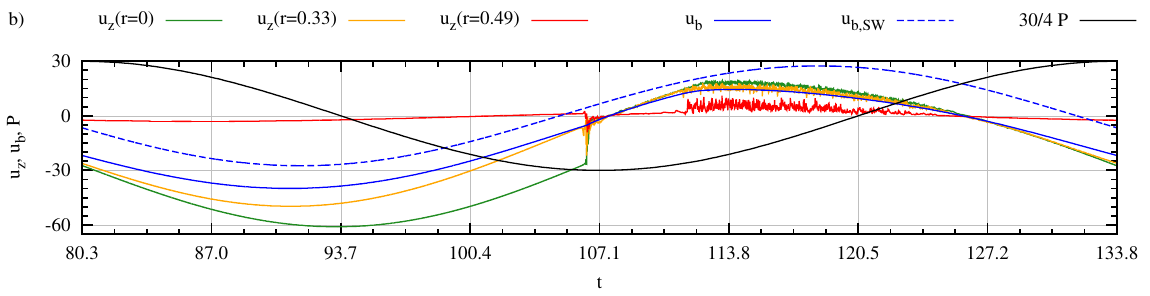}
\includegraphics[scale=1.00]{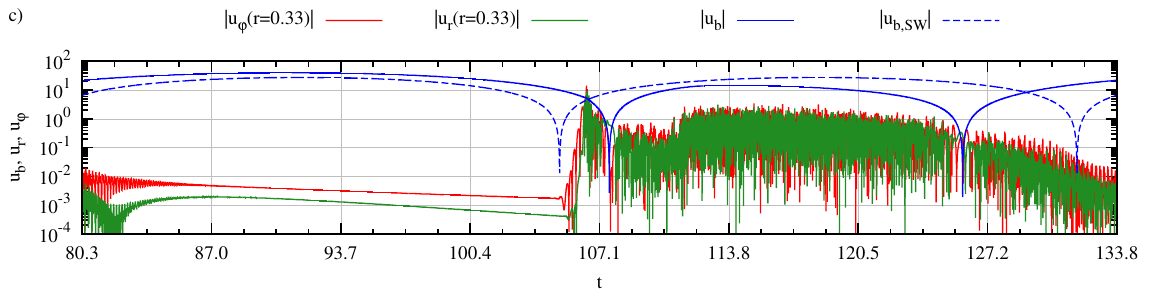}
\caption[Time series for $\Womersley=\num{6.5}$]
{Time series of all velocity components, the bulk flow, and the
wall-shear stress as predicted by DNS in a pipe domain of length $L^{+}=\num{7200}$.
The flow is driven by a prescribed pressure gradient $P(t)$ for $\Womersley=\num{6.5}$.
Analytical predictions according to laminar theory (SW) are shown as reference.}
\label{fig:timeLinesAsymmetric}
\end{figure}

%
The temporal evolution of all three velocity components recorded at different
radial locations is shown in fig.~\ref{fig:timeLinesAsymmetric} for an
oscillatory pipe flow at $\Womersley=\num{6.5}$.
%
%
There is a strong asymmetry between half-cycles of positive and negative bulk
flow values, even though the prescribed driving pressure gradient is symmetric.
The velocity field is uniform and of laminar nature during every other
half-cycle with $u_{\text{b}}<\num{0}$, e.g. $\num{72.3}<t<\num{107.6}$.
At the end of those half-cycles there is always a sudden breakdown of the
laminar state.
Each subsequent half-cycle with $u_{\text{b}}>\num{0}$ , e.g.
$\num{107.6}<t<\num{125.9}$, is then characterised by strong velocity
fluctuations and a turbulent flow field.
\par
The recurring of such events like the RV and the laminar breakdown is highly
periodic, already after the first DC phase.
For example, the length of the laminar half-cycles is
$T_{\text{lam}}=\num{35.2}\pm\num{0.1}$ and the length of the turbulent ones is
$T_{\text{turb}}=\num{18.3}\pm\num{0.1}$.
Thus, we exclude the possibility that the found asymmetric oscillation is only a
transient state which will disappear in advantage of a symmetric one.
\par
\begin{figure}[htbp]
\footnotesize{a) $t=\num{83.4}$}\\
\includegraphics[width=1.00\textwidth]{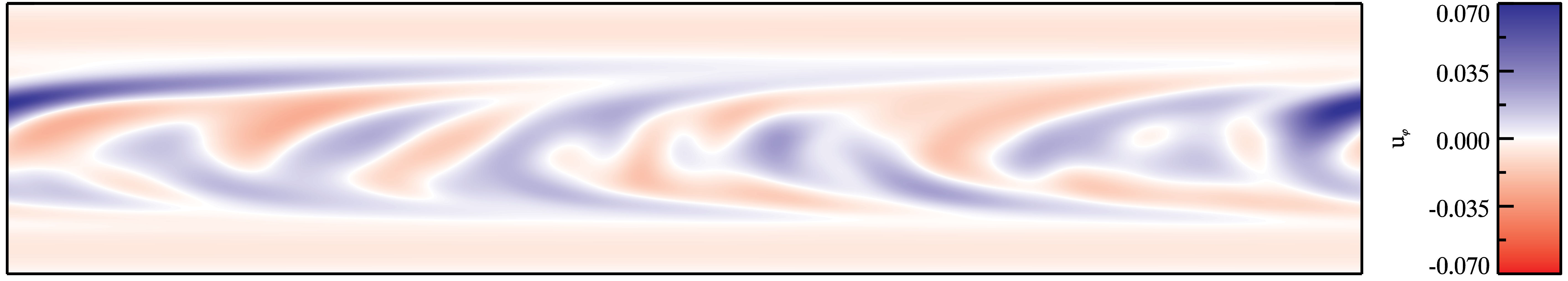}
\footnotesize{b) $t=\num{114.2}$}\\
\includegraphics[width=1.00\textwidth]{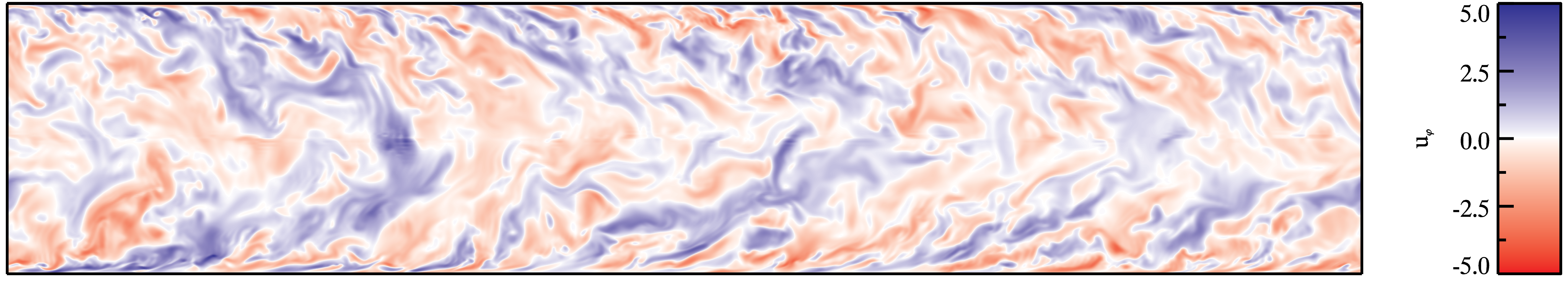}
\caption[Velocity field for $\Womersley=\num{6.5}$]
{Colour coded contour plots of the azimuthal velocity component $u_{\varphi}$,
as predicted by DNS of oscillatory flow at $\Womersley=\num{6.5}$ in a pipe
domain of length $L^{+}=7200$.}
\label{fig:contourPlotsAsymmetric}
\end{figure}

In the DC phase of the laminar half-cycle a phase lag develops in the velocity
field between the near-wall and the central region due to the interplay of
viscous and inertial forces.
This can be seen in fig. \ref{fig:timeLinesAsymmetric}b) at
$t\approx\num{100.4}$, where the fluid in the vicinity of the wall is moving in
the positive $z$ direction already, whereas the fluid further away from the wall
is still moving in the negative direction.
Such a phase lag is typical for SW flow at $\Womersley>\num{1}$ and leads to
maxima and inflection points in the velocity distribution, cf. fig.
\ref{fig:velocityProfilesEckmann}.
The latter also represent a source of instabilities, which are most likely
triggered by small residual disturbances remaining in the secondary components
($u_{\varphi}$, $u_{r}$) of the velocity field since the last turbulent
half-cycle.
Fig. \ref{fig:contourPlotsAsymmetric}a) shows the structure of such disturbances
in the azimuthal velocity component during the laminar AC phase.
%
%
Fig. \ref{fig:timeLinesAsymmetric}c) highlights the longsome decay of the
kinetic energy content in $u_{\varphi}$ and $u_{z}$ in the course of the laminar
half-cycle.
Nevertheless, right before the event of laminar breakdown the secondary velocity
components are still $\mathcal{O}\num{e-2}$.
They rapidly grow by three orders of magnitude within two or three convective
time units and lead to a highly turbulent flow field, e.g.
$\num{104.6}<t<\num{107.1}$.
\par
The following positive half-cycle is characterised by chaotic velocity
fluctuations during early AC and fully-developed turbulence during the
rest of the half-cycle, as shown in fig. \ref{fig:contourPlotsAsymmetric}b).
Due to the onset of turbulence the wall-shear stress increases, see fig.
\ref{fig:timeLinesAsymmetric}a), and thus the turbulent flow only accelerates up
to $u_{\text{b}}=\num{14.6}\pm\num{0.2}$.
This corresponds to $\Reynolds=\num{21000}$ which is of course much lower than
the predicted value from laminar theory, see fig. \ref{fig:woReRange}.
Fig. \ref{fig:timeLinesAsymmetric} further reveals that the turbulence intensity
is decreasing in accordance with the decelerating bulk flow.
Nevertheless, the velocity fluctuations are still $\mathcal{O}\num{e-1}$ at RV.
The wall-normal momentum transport due to turbulent mixing prevents the
formation of near-wall velocity maxima, inflection points and co-axial counter
flows.
In fact, the velocity distribution during RV is rather uniform; i.e.
$u_{z}\approx\num{0}$ for all $r$.
The initial condition for any laminar half-cycle is therefore very similar
to the classic scenario of a start-up flow in a pipe. 
Missing inflection points and the stabilising effect of acceleration promote
the ongoing laminarisation.
The initially resting fluid and the much lower wall-shear stress allow the
laminar flow to accelerate up to $u_{\text{b}}=\num{39.8}\pm\num{0.1}$.
This corresponds to $\Reynolds=\num{57370}$ which is much higher compared to SW
theory, see fig.~\ref{fig:woReRange}.
Therefore, the laminar as well as the turbulent half-cycle is characterised by
a $\Reynolds_{\delta}$ much higher than the critical value
$\Reynolds_{\delta,\text{crit}}\approx\num{550}$ for the onset of turbulence
found experimentally by Hino et al. \cite{hino1976}, Eckmann \& Grotberg
\cite{eckmann1991}, and Zhao \& Cheng \cite{zhao1996}.
%
%
%

%
%
%
\section{Conclusion}
\label{sec:conclusion}
%
We report on strong asymmetries in an oscillatory pipe flow at
$\Womersley=\num{6.5}$ and $\Reynolds_{\tau}=\num{1440}$ investigated by means
of DNS using a prescribed pressure force driving the flow.
Despite the problem being symmetric in terms of its boundary conditions, the
resulting flow is turbulent during half-cycles of positive bulk flow values and
laminar during half-cycles of negative bulk flow values.
%
%
The turbulent phase is shorter, the peak flow rate is smaller, and the
following bulk flow reversal happens earlier, when compared to predictions
from SW theory for laminar oscillating pipe flow.
Contrarily, the laminar phase is longer, the reached peak flow rate is
larger, and the following RV happens later as expected from SW theory.
As a result, the oscillating pipe flow develops asymmetric transport and mixing
properties.
\par
The used numerical method and the spatio-temporal resolution was validated for
the problem at hand by demonstrating excellent agreement of our DNS results with
experimental data and analytical predictions from literature for the two
limiting cases of a non-oscillating but turbulent pipe flow
($\Womersley=\num{0}$) and an oscillating but laminar pipe flow
($\Womersley=\num{26}$).
For an oscillating and turbulent pipe flow ($\Womersley=\num{13}$) we found very
good agreement with qualitative descriptions of the flow field observed in
experiments.
For the non-oscillating case ($\Womersley=\num{0}$) we have shown earlier
\cite{feldmann2012}, that $L^+=\num{1800}$ is a sufficiently long pipe domain
and the coarser grid is sufficiently fine to capture most of the turbulent
scales.
For $\Womersley=\num{13}$, results from DNS employing three different pipe
domains with $L^+=\num{1700}$ up to $L^+=\num{5100}$ qualitatively show the same
behaviour of the oscillating flow field.
For $\Womersley=\num{6.5}$ we performed additional DNS in even longer domains
to make sure that the asymmetry is not an artefact caused by the finite pipe
length.
Results from DNS using $L^+=\num{1800}$ and $L^+=\num{7200}$ show the same
behaviour of the oscillating flow field.
A more detailed study of the influence of the length of the computation pipe
domain in oscillatory turbulent flow is the subject of our ongoing work
\cite{feldmann2015}.
\par
To our knowledge, this asymmetric behaviour was never found in experiments on
oscillatory pipe flow published in literature.
The major difference between our DNS and all experimental set-ups is the
forcing mechanism which drives the oscillatory flow through the pipe.
In pipe flow experiments it is most convenient to drive the flow by a
reciprocating piston; i.e. the flow rate is prescribed and thus always follows
a sine function.
The pressure gradient across the pipe is a result and not necessarily symmetric.
Contrarily, in our DNS the pressure gradient is prescribed to follow a sine
function due to convenience.
The temporal evolution of the bulk flow is therefore a result and not
necessarily symmetric.
The asymmetry is introduced by the reduced length of the turbulent
half-cycle with $T_{\text{lam}}/T_{\text{turb}}=1.9$.
Thus the driving force is rather small at this preterm RV; but further
increasing.
On the other hand the length of the laminar half-cycle is enlarged
and therefore the driving force has already reached its peak value at 
this delayed RV.
For completely laminar and fully developed oscillatory pipe flows a
difference in the driving mechanism only results in a phase shift.
But for non-laminar oscillating pipe flows, this difference can lead
to strong asymmetries either in the temporal evolution of the
resulting flow or in the temporal evolution of the resulting pressure drop.
Thus, the difference in boundary conditions is a crucial point and has to be
considered for future numerical studies on oscillatory pipe flow. 
%
%
%

%
\bibliographystyle{closing/spmpsci.bst}
\bibliography{closing/stab2014.bib}
%
\end{document}